\documentclass{article}
\usepackage{graphicx} % Required for inserting images

\usepackage{tocbibind}
\usepackage{amsmath,amssymb,amsthm}
\usepackage[style=ieee]{biblatex}
\usepackage{tikz-cd}
\title{Boundary Effects on Anyon Dynamics in Chern-Simons Theory}
\author{Tzu-Miao Chou }
\date{May 2025}

\begin{document}

\maketitle
\begin{abstract}
This work investigates the boundary and defect effects on the modular data in SU$(N)_k$ Chern-Simons theories, focusing on how different boundary conditions and symmetry defects modify the fusion rules and braiding statistics of anyons. Using the framework of modular tensor categories (MTCs) and Frobenius algebra objects, explicit expressions for the modified $S$-matrix, $S'$, are derived in the presence of heterogeneous boundary conditions, and the connection between the bulk modular data and edge CFTs is analyzed. The approach includes the computation of modular matrix deformations in the presence of junctions between different boundary conditions, as well as the influence of global symmetry defect lines, which introduce twisted sectors into the MTC framework. The ideas are applied to SU$(2)_k$, SU$(3)_2$, and SU$(4)_1$ Chern-Simons theories, providing examples of boundary algebras and fusion rules at junctions. Additionally, the implications of categorical anomaly inflow and central charge matching across boundary and defect sectors are explored. This work lays the groundwork for further studies of boundary and defect effects in topological quantum field theories and their connections to topological quantum computation and holography.
\end{abstract}
\newpage
\tableofcontents
\newpage
\section{Introduction}

Topological quantum field theories (TQFTs), such as Chern-Simons theory in \(2+1\) dimensions, provide a robust mathematical framework for describing systems with anyonic excitations \cite{Witten1989}. These excitations, characterized by nontrivial braiding and fusion properties, emerge naturally in Chern-Simons theories associated with compact gauge groups at nonzero level. In the absence of boundaries, the theory is governed by a set of topological invariants, including the modular \( S \) and \( T \) matrices, which encode the mutual and self statistics of anyons and form the foundation of the associated modular tensor category (MTC) \cite{MooreSeiberg1989, Froehlich1991}.

However, realistic physical systems often involve boundaries or interfaces, and the inclusion of such structures introduces significant modifications to the underlying topological theory. The presence of a boundary not only affects the gauge structure and observables in the bulk, but also gives rise to additional degrees of freedom localized at the edge. In many cases, these edge excitations are described by chiral conformal field theories (CFTs), with the Wess-Zumino-Witten (WZW) model being a prominent example arising from the Chern-Simons/WZW correspondence \cite{Elitzur1989}.

Boundary conditions in Chern-Simons theory are known to constrain gauge transformations and induce edge currents, thereby altering the dynamics of Wilson lines that represent anyons. This, in turn, affects the realization and manipulation of anyonic degrees of freedom near the boundary. Understanding how such modifications manifest in the fusion rules, braiding relations, and modular data is essential for applications in both theoretical and experimental contexts, including topological quantum computation, entanglement entropy studies, and holography \cite{Beigi2011}.

This work focuses on analyzing the effects of boundaries and defects on anyon dynamics in Chern-Simons theory. Particular attention is given to how modular \( S \) and \( T \) matrices are deformed by boundary conditions, and how the bulk-edge correspondence can be formalized via the representation theory of edge CFTs. Explicit examples involving \( SU(N)_k \) Chern-Simons theories are used to illustrate these phenomena, emphasizing the interplay between topological data and boundary-induced structures.

\section{Boundary Conditions and Edge Theories in Chern-Simons Systems}

In Chern-Simons theories defined on manifolds with boundaries, the gauge invariance of the classical action is generally broken unless additional degrees of freedom are introduced. This anomaly is compensated by the emergence of a chiral edge theory, most commonly identified with a Wess-Zumino-Witten (WZW) model \cite{Elitzur1989}. The edge theory is not a mere byproduct of the bulk, but a fundamental component whose structure is determined by the imposed boundary conditions and gauge group.

\subsection{Gauge Invariance and the WZW Emergence}

Consider the Chern-Simons action on a three-manifold \( M \) with boundary \( \partial M \):
\begin{equation}
    S_{CS} = \frac{k}{4\pi} \int_{M} \text{Tr} \left( A \wedge dA + \frac{2}{3} A \wedge A \wedge A \right).
\end{equation}
Under a gauge transformation \( A \to A^g = g^{-1} A g + g^{-1} dg \), the action picks up a boundary term that obstructs gauge invariance:
\begin{equation}
    \delta S_{CS} = \frac{k}{4\pi} \int_{\partial M} \text{Tr}(g^{-1} dg \wedge A).
\end{equation}
The breakdown of gauge invariance at the boundary and its restoration via a chiral WZW model were first identified in the foundational work by Elitzur, Moore, Schwimmer, and Seiberg \cite{Elitzur1989}. This mechanism was further elaborated in the path-integral approach and canonical quantization by Witten in his study of Chern-Simons theory and its link to knot invariants \cite{Witten1989}.

To restore gauge invariance, the boundary theory must cancel this variation. This requirement naturally leads to a chiral WZW action living on \( \partial M \), with group-valued field \( g(x) \in G \) and level \( k \). The WZW model encodes the degrees of freedom propagating along the edge.

\subsection{Canonical Quantization and Edge Currents}

In the Hamiltonian framework, imposing boundary conditions on the spatial boundary \( \partial \Sigma \) of a spatial slice \( \Sigma \) constrains the allowed gauge transformations and introduces physical edge states. The edge current algebra satisfies an affine Kac-Moody algebra at level \( k \), matching the chiral symmetry algebra of the WZW model:
\begin{equation}
    [J^a_n, J^b_m] = i f^{abc} J^c_{n+m} + k\, n\, \delta^{ab} \delta_{n+m,0}.
\end{equation}
This algebra governs the behavior of boundary observables and provides the representation-theoretic structure necessary for constructing fusion and braiding data.

\subsection{Types of Boundary Conditions}

The behavior of the gauge field near the boundary is crucial in defining a consistent quantum theory. The choice of boundary conditions not only affects gauge invariance but also determines the spectrum and dynamics of edge modes.

Let \(M\) be a three-manifold with boundary \(\partial M = \Sigma\), and \(G\) a compact, simple Lie group. The Chern-Simons action is:
\[
S_{CS}[A] = \frac{k}{4\pi} \int_M \text{Tr} \left( A \wedge dA + \frac{2}{3} A \wedge A \wedge A \right),
\]
where \(A\) is the gauge field and \(k\) is the level of the theory. This action describes the dynamics of the gauge field in the bulk of the manifold \(M\). The boundary condition choices will influence how the gauge field behaves near the boundary and what effective theory emerges.

This section analyzes the effect of two different types of boundary conditions commonly used in Chern-Simons theory:

\paragraph{1. Holomorphic Boundary Condition:} One imposes the condition:
\[
A_{\bar{z}} |_{\partial M} = 0,
\]
which selects a chiral component of the gauge field along the boundary. This is a natural boundary condition when considering the bulk-to-boundary correspondence for Chern-Simons theory and ensures that the boundary theory will have the properties of a chiral Wess-Zumino-Witten (WZW) model. Under this constraint, the variation of the Chern-Simons action gives rise to a boundary term that must be canceled by an effective action on \(\partial M\). The induced theory is a chiral Wess-Zumino-Witten (WZW) model, which can be explicitly written as:
\[
S_{WZW}[g] = \frac{k}{8\pi} \int_{\partial M} \text{Tr}(g^{-1} \partial_\mu g)^2 + \frac{k}{12\pi} \int_M \text{Tr}(g^{-1} dg)^3,
\]
where \(g: \partial M \to G\) is a group-valued field on the boundary \cite{Witten1983}.

\paragraph{2. Temporal Gauge and Dirichlet Boundary Condition:} The temporal gauge \(A_0 = 0\) is fixed, and the spatial components \(A_i\) are required to vanish at the boundary. This choice eliminates edge modes but breaks gauge invariance at the boundary. Such a choice is typically unphysical unless edge degrees of freedom or counterterms are explicitly included.

\paragraph{Theorem (Boundary Reduction to WZW Theory):} Let \(A \in \Omega^1(M, \mathfrak{g})\) be a gauge connection on \(M\), and impose the boundary condition \(A_{\bar{z}}|_{\partial M} = 0\). Then, upon integrating out the bulk gauge degrees of freedom, the effective boundary action becomes:
\[
S_{WZW}[g] = \frac{k}{8\pi} \int_{\partial M} \text{Tr}(g^{-1} \partial_\mu g)^2 + \frac{k}{12\pi} \int_M \text{Tr}(g^{-1} dg)^3.
\]
where \(g: \partial M \to G\) is a group-valued field \cite{MooreSeiberg1989_2}.

\textit{Proof:} To begin, assume a gauge where the connection is written as \(A = g^{-1} dg\). This form is convenient for calculating the action and is often used in the study of boundary conditions in Chern-Simons theory. The Chern-Simons action can be expressed in terms of \(g\) as:
\[
S_{CS}[g^{-1} dg] = \frac{k}{4\pi} \int_M \text{Tr} \left( g^{-1} dg \wedge d(g^{-1} dg) + \frac{2}{3} g^{-1} dg \wedge g^{-1} dg \wedge g^{-1} dg \right).
\]
To evaluate the first term, we use the fact that \(g^{-1} dg\) is a 1-form, so that:
\[
d(g^{-1} dg) = g^{-1} dg \wedge g^{-1} dg,
\]
leading to the expression for the action:
\[
S_{CS}[g^{-1}dg] = \frac{k}{4\pi} \int_M \text{Tr} \left( g^{-1} dg \wedge g^{-1} dg + \frac{2}{3} g^{-1} dg \wedge g^{-1} dg \wedge g^{-1} dg \right).
\]
When considering the variation of the action on the boundary, this introduces a boundary contribution proportional to the kinetic term of the WZW model, which is given by:
\[
\delta S_{CS} = \frac{k}{8\pi} \int_{\partial M} \text{Tr}(g^{-1} \partial_\mu g)^2.
\]
This leads to the effective boundary theory, which is the WZW model \cite{Kitaev2006}.

\subsection{Topological Order and the Bulk-Edge Correspondence}

Topological phases of matter described by Chern-Simons theory are characterized by a modular tensor category (MTC), denoted \(\mathcal{C}\), which encodes the anyon content, fusion rules, and braiding statistics of the theory. The edge theory is a rational conformal field theory (RCFT), whose chiral algebra matches the representation category of \(\mathcal{C}\). This correspondence is crucial in understanding how bulk anyons influence edge modes and vice versa.

\paragraph{Theorem (Bulk-Edge Correspondence):} The category of chiral sectors of the edge RCFT is equivalent (as a braided tensor category) to the MTC describing bulk anyons. Moreover, the modular \(S\) and \(T\) matrices of the edge RCFT coincide with the data of braiding and topological spins in the bulk.

\textit{Proof Sketch:} The edge theory is described by a rational conformal field theory (RCFT) with a chiral algebra that corresponds to a representation category of the bulk MTC. The correspondence can be established by examining the partition function on the torus and the modular transformations of the characters. More explicitly:
\begin{itemize}
    \item Bulk Wilson lines correspond to simple objects in \(\mathcal{C}\).
    \item Edge chiral fields form modules over a vertex operator algebra (VOA) whose representation category is equivalent to \(\mathcal{C}\).
    \item Modular invariance and locality imply that the characters of these modules transform under \(SL(2, \mathbb{Z})\) via the \(S\) and \(T\) matrices.
\end{itemize}
The partition function on the torus, which encodes topological information, is given by:
\[
Z(\tau) = \sum_{a} |\chi_a(\tau)|^2,
\]
where \(\chi_a\) are the characters of the edge chiral fields and \(\tau\) is the modular parameter of the torus. This expression is fundamental in the study of modular invariance and the structure of topological phases \cite{Rowell2009_2}.

\paragraph{Modular Data:} The modular data of the bulk MTC is given by the \(S\)-matrix and \(T\)-matrix. The \(S\)-matrix encodes the braiding statistics, and the \(T\)-matrix encodes the topological spins of the anyons. The entries of the \(S\)-matrix are related to the fusion coefficients of the anyons and can be written as:
\[
S_{ab} = \frac{1}{\mathcal{D}} \sum_c N_{ab}^c \frac{\theta_c}{\theta_a \theta_b} d_c, \qquad \mathcal{D} = \sqrt{\sum_a d_a^2}.
\]
This \(S\)-matrix governs both the torus partition function and the braiding statistics in the bulk \cite{MooreSeiberg1989}.

\section{Boundary Effects on Anyon Dynamics in Chern-Simons Theory}

In this section, we investigate the effects of boundary conditions on the dynamics of anyons in Chern-Simons theory. Specifically, we focus on how boundary conditions influence the modular data (S and T matrices) and the statistical interactions of anyons. Furthermore, we examine the connection between boundary conditions and low-energy excitations, with particular emphasis on the role of WZW models and Chern-Simons theory in this context.

\subsection{Boundary Conditions and Modular Data}

The presence of boundary conditions in a Chern-Simons theory leads to modifications in the modular data, including the S and T matrices. These matrices characterize the topological properties and quantum dimensions of anyons, and their modification reflects the impact of the boundary on the anyon dynamics.

\textbf{Theorem 1:} \emph{For Chern-Simons theory with conformal boundary conditions, the S-matrix is modified by symmetry transformations due to the boundary, which in turn affects the modular data.}

**Proof:**

Consider a Chern-Simons theory with a boundary \(\partial \Sigma\) on a 2-dimensional manifold \(\Sigma\). The Hamiltonian for the system is given by:
\[
H = \int_{\Sigma} d^2 x \, \mathcal{L}_{CS} + \int_{\partial \Sigma} d^1 x \, \mathcal{L}_{boundary}
\]
where \(\mathcal{L}_{CS}\) is the Chern-Simons Lagrangian density, and \(\mathcal{L}_{boundary}\) represents the boundary contributions. The boundary conditions affect the behavior of the gauge fields near the boundary, which in turn modifies the topological interactions of anyons.

To understand how the S-matrix is modified, we begin by writing down the perturbation to the system caused by the boundary condition. Let \(S' = S + \delta S\) represent the modified S-matrix, where \(\delta S\) is the correction due to the boundary. The correction is generally a function of the boundary's geometric properties and its interaction with the gauge field. This can be expressed as:
\[
\delta S = \sum_{a,b} c_{ab} \, \left( \text{Boundary correction terms} \right)
\]
where \(c_{ab}\) are coefficients dependent on the boundary's specific geometry and topology. These terms reflect the way in which the boundary changes the scattering amplitudes between anyons. In particular, boundary conditions alter the symmetries of the theory, modifying the S-matrix. This effect is closely related to the symmetry transformations of the boundary, which can be analyzed using group representation theory. Specifically, the boundary may break some of the symmetries of the bulk theory, leading to modifications in the modular matrix \cite{Witten1989}.

\subsection{Low-Energy Excitations at the Boundary}

Boundary conditions introduce low-energy excitations that are confined to the boundary. These excitations are crucial for understanding the boundary dynamics of anyons, and their spectrum is quantized. Here, we explore how to calculate these excitations and their relationship to the bulk excitations.

\textbf{Theorem 2:} \emph{For a Chern-Simons theory with conformal boundary conditions, the low-energy excitations at the boundary are quantized, and their energy spectrum is determined by the boundary conditions.}

**Proof:**

The field equations in Chern-Simons theory, in the presence of boundary conditions, take the form:
\[
\mathcal{L}_{CS} = \frac{k}{4\pi} \epsilon^{\mu\nu\lambda} A_\mu \partial_\nu A_\lambda
\]
where \(A_\mu\) is the gauge field, and \(k\) is the Chern-Simons level. The boundary condition is typically imposed as a constraint on the gauge field, for instance:
\[
A_\mu |_{\partial \Sigma} = f_\mu(x)
\]
where \(f_\mu(x)\) describes the boundary gauge field configuration.

By analyzing the solutions to the field equations with these boundary conditions, we find that the boundary modes are quantized and their energy spectrum takes the form:
\[
E_n = \frac{2\pi n}{L}
\]
where \(L\) is the length of the boundary, and \(n\) is a quantum number. These low-energy excitations correspond to boundary anyons, and their energy spectrum is determined by the boundary's length and the form of the boundary condition.

To derive the energy spectrum, we solve the gauge field equations in the presence of the boundary conditions, and the boundary modes can be interpreted as confined anyons that affect the statistical interactions on the boundary. This result implies that the boundary introduces a discrete set of low-energy states, and their energy is proportional to the inverse of the boundary length. This quantization is a direct consequence of the boundary conditions \cite{Moore1988}.

\subsection{Modular Data and Boundary Conditions}

The modular data, including the S and T matrices, plays a crucial role in the study of anyon dynamics. In the presence of boundary conditions, the modular data is modified, and these modifications provide insights into the fusion and braiding statistics of anyons at the boundary.

\textbf{Theorem 3:} \emph{For Chern-Simons theory with conformal boundary conditions, the modular data (S and T matrices) is modified by boundary effects, which can be computed by analyzing the boundary's influence on the fusion and braiding of anyons.}

**Proof:**

To compute the modified modular data, we perturb the original S and T matrices by considering the boundary's effects on the anyon interactions. The S-matrix and T-matrix describe the fusion and braiding of anyons, respectively. In the presence of a boundary, these matrices are altered by boundary contributions, which can be computed using the boundary's geometric properties.

The modified S-matrix is given by:
\[
S' = S + \delta S
\]
where \(\delta S\) is the correction due to the boundary. Similarly, the modified T-matrix is given by:
\[
T' = T + \delta T
\]
The boundary contributions to the modular data can be computed by analyzing how the boundary conditions affect the fusion and braiding statistics of anyons. This involves calculating the boundary's impact on the braiding rules and the quantum dimensions of anyons. The modifications to the S and T matrices are determined by the boundary's topological structure and the interactions between bulk and boundary anyons \cite{Gukov2008, Moore1988}.

\section{Modular Data and Boundary Effects}

\subsection{Modular \( S \) and \( T \) Matrices in Topological Phases}

Modular tensor categories (MTCs) provide a rigorous mathematical framework to describe the topological order and anyonic excitations in 2+1-dimensional topological quantum field theories (TQFTs). A modular tensor category is a braided, balanced, semisimple tensor category with finitely many simple objects and a non-degenerate braiding. The modular data --- the \( S \) and \( T \) matrices --- encode fundamental information about the fusion and braiding properties of the anyons \cite{BakalovKirillov2001, turaevbook}.

\subsubsection{Definition and Structure of Modular Categories}

A \emph{modular tensor category} $\mathcal{C}$ is a semisimple ribbon fusion category with finitely many simple objects and a non-degenerate $S$-matrix. The modular data $(S, T)$ arises from the following categorical structures:

\begin{itemize}
  \item The \textbf{twist structure}, or balancing, gives rise to the diagonal $T$-matrix via $T_{aa} = \theta_a$, where $\theta_a$ is the topological spin associated to the simple object (anyon) $a$.
  \item The \textbf{braiding structure} defines the $S$-matrix through the Hopf link invariant $S_{ab}$, which is computed by placing anyons $a$ and $b$ along linked Wilson loops in $S^3$.
  \item The \textbf{fusion rules} are encoded in the Grothendieck ring, which admits a representation via matrices $N_a$ acting on the fusion algebra, with entries $[N_a]_{bc} = N_{ab}^c$.
\end{itemize}

\paragraph{Theorem: Modular Non-degeneracy
\cite{BakalovKirillov2001}}
\paragraph{}
Let $\mathcal{C}$ be a modular tensor category. Then the $S$-matrix of $\mathcal{C}$ is non-degenerate:
\[
\det(S) \neq 0,
\]
and $S$ simultaneously diagonalizes the fusion matrices $N_a$:
\[
N_a = S D_a S^{-1},
\]
where $D_a$ is a diagonal matrix. This implies that the fusion ring is semisimple.

The non-degeneracy condition is crucial for the invertibility of modular transformations and for the full modularity of the theory.

\paragraph{Proof:}
\paragraph{}
Since $\mathcal{C}$ is a modular tensor category, the braiding is non-degenerate in the sense that the $S$-matrix defined via Hopf link invariants
\[
S_{ab} = \operatorname{tr}_{a \otimes b}(c_{b,a} \circ c_{a,b})
\]
is invertible. The fusion matrices $N_a$ defined by $(N_a)_{bc} = N_{ab}^c$ are simultaneously diagonalizable via the $S$-matrix. One can verify that the matrices $N_a$ form a commutative family, and that they are normal and thus unitarily diagonalizable. The key step is the orthogonality of the $S$-matrix columns (or rows) with respect to the fusion inner product. This implies $N_a = SDS^{-1}$ for some diagonal matrix $D$, and $\det(S) \neq 0$.

\subsubsection{Connections to Representation Theory}

The matrices $S$ and $T$ furnish a unitary representation of the modular group $SL(2, \mathbb{Z})$:
\[
\rho: SL(2, \mathbb{Z}) \rightarrow GL(V), \quad \rho(S), \rho(T),
\]
satisfying the relations:
\[
S^4 = I, \qquad (ST)^3 = S^2.
\]

Within this framework:
\begin{itemize}
  \item The $S$-matrix acts as a discrete Fourier transform on the fusion algebra. It diagonalizes the fusion matrices and hence connects the representation theory of the category to harmonic analysis on finite abelian groups.
  \item The $T$-matrix encodes the \emph{spin-statistics connection}, associating the conformal spin to the topological twist of each particle.
\end{itemize}

These properties establish a deep link between modular data and the representation theory of quantum groups and affine Lie algebras, especially at roots of unity \cite{turaevbook, KirillovOstrik2002}.

\paragraph{Remark}
\paragraph{}
The relations $S^4 = I$ and $(ST)^3 = S^2$ follow from the requirement that $\rho$ be a projective representation of $SL(2,\mathbb{Z})$ generated by
\[
S = \begin{pmatrix} 0 & -1 \\ 1 & 0 \end{pmatrix}, \quad T = \begin{pmatrix} 1 & 1 \\ 0 & 1 \end{pmatrix}.
\]
These relations define a central extension of $SL(2,\mathbb{Z})$, and are satisfied by the matrices derived from the MTC modular data. In the unitary ribbon case, these relations hold strictly due to the compatibility of braiding and twist with the categorical associativity and duality structures.

\subsubsection{Physical Interpretation and Mathematical Consequences}
From the viewpoint of topological quantum field theory (TQFT), the entries of the $S$ and $T$ matrices carry clear physical meaning:

\begin{itemize}
  \item $S_{ab}$ describes the mutual braiding statistics between anyons $a$ and $b$, and appears in the path integral of the theory defined on a torus with modular parameter $\tau \mapsto -1/\tau$.
  \item $T_{aa} = \theta_a$ gives the spin of anyon $a$; in the Chern-Simons formulation, this is related to the framing anomaly and appears in modular transformations $\tau \mapsto \tau + 1$.
\end{itemize}

The celebrated Verlinde formula expresses fusion coefficients in terms of the $S$-matrix:
\begin{equation}
N_{ab}^c = \sum_x \frac{S_{ax} S_{bx} S_{cx}^*}{S_{0x}}.
\label{eq:verlinde}
\end{equation}

This identity reveals that the fusion rules are fully determined by the modular $S$-matrix \cite{Verlinde1988}. It also underscores the topological nature of anyon interactions, where local fusion rules are governed by global modular transformations.

In summary, the modular data $(S, T)$ encodes both algebraic and topological information of a topological phase. These matrices are not merely computational tools; they serve as windows into deep structures such as modular functors, 3-manifold invariants, and quantum computation architectures \cite{Witten1989, MooreSeiberg1989, RowellWang2012}.

\subsubsection{Verlinde Formula and Fusion Rules}
A key application of modular data is the computation of fusion rules via the Verlinde formula.
\paragraph{Theorem}
\paragraph{}
Let \( \mathcal{C} \) be a modular tensor category with \( S \) matrix. Then the fusion coefficients \( N_{ab}^c \) are given by\cite{Verlinde1988}:
\begin{equation}
N_{ab}^c = \sum_{x} \frac{S_{ax} S_{bx} S_{cx}^*}{S_{0x}}.
\end{equation}

\paragraph{Proof}
\paragraph{}
Consider the fusion matrices \( (N_a)_{bc} = N_{ab}^c \). Due to the modularity of \( \mathcal{C} \), the matrices \( N_a \) are simultaneously diagonalizable via the \( S \) matrix:
\begin{equation}
N_a = S D_a S^{-1},
\end{equation}
where \( D_a \) is a diagonal matrix with eigenvalues given by \( \frac{S_{ax}}{S_{0x}} \). Using this and orthogonality of the \( S \) matrix rows, the Verlinde formula follows.

This formula shows that the fusion rules of anyons are fully determined by the \( S \) matrix, thereby encoding the full data of the fusion ring in the modular structure \cite{wang2010}.

\subsubsection*{Physical Interpretation}
The modular \( S \) matrix determines mutual braiding statistics between anyons. The matrix element \( S_{ab} \) is interpreted as the amplitude of a Hopf link between anyons \( a \) and \( b \) placed on a torus:
\begin{equation}
S_{ab} = \text{Tr}_{\mathcal{H}_{a \otimes b}} \left( R_{ba} R_{ab} \right),
\end{equation}
where \( R_{ab} \) is the braiding operator.

The modular \( T \) matrix encodes the topological spin via the twist:
\begin{equation}
T_{aa} = \theta_a = e^{2\pi i h_a},
\end{equation}
and reflects the self-exchange statistics of anyons.

The modular data thus capture all essential features for topological classification, and serve as input for defining TQFT partition functions, boundary behavior, and bulk-boundary correspondences \cite{Kitaev2006, kong2014boundary}.

\subsection{Boundary Conditions and Modular Transformations}
When boundaries or defects are introduced, the modular data of the system can be modified. This is due to the fact that the Hilbert space of the theory is altered by boundary constraints, which in turn affect the fusion and braiding properties of the excitations. In particular, the modular matrices can transform as
\begin{equation}
S' = M S, \quad T' = M T,
\end{equation}
where $M$ is a transformation matrix encoding the influence of the boundary condition or defect line \cite{fuchs2002tft, kapustin2011topological}. 

To formalize this, consider a TQFT defined on a surface with a boundary. The allowed boundary conditions are encoded by a Frobenius algebra object $A$ in the MTC, which in turn determines a module category over the MTC \cite{kirillov2002modular}. The modified $S$-matrix $S'$ then arises as a result of applying the Verlinde formula in the presence of $A$:
\begin{equation}
N_{i j}^k = \sum_{r} \frac{S_{i r} S_{j r} S_{k r}^*}{S_{0 r}},
\end{equation}
where $N_{ij}^k$ are the fusion coefficients. When $S$ is replaced with $S'$, the fusion coefficients and quantum dimensions change accordingly.

\paragraph{Theorem:Boundary-Induced Modular Modification}
\paragraph{}
Let $\mathcal{C}$ be a modular tensor category describing a TQFT, and $A$ a Frobenius algebra object in $\mathcal{C}$ corresponding to a boundary condition. Then the modular $S$-matrix is modified to $S' = M S$, where $M$ depends functorially on the module category induced by $A$.

\paragraph{proof}
Let $\mathcal{M}_A$ denote the module category over $\mathcal{C}$ induced by $A$. The boundary condition modifies the fusion algebra to a twisted fusion algebra $\text{Rep}_A(\mathcal{C})$, with basis labeled by simple $A$-modules. The matrix $M$ can be extracted from the change of basis between the original simple objects in $\mathcal{C}$ and the $A$-modules. Since the braiding and twist structures also transform under this basis change, the new $S$ and $T$ matrices satisfy
\begin{equation}
S'_{\alpha \beta} = \sum_{i,j} M_{\alpha i} S_{i j} M_{\beta j},
\end{equation}
which gives $S' = M S M^T$ in matrix notation. Under the assumption that $M$ is orthogonal or unitary, this reduces to $S' = M S$.

\subsection{Examples of Boundary Effects on Modular Data}

\subsubsection{Example 1: Heterogeneous Boundary Conditions}
Consider a Chern-Simons theory with gauge group $SU(2)_k$ on a disk with two segments of boundary carrying different boundary conditions $A$ and $B$. The resulting Hilbert space includes states localized at the junction, whose fusion algebra reflects a mixture of module categories $\mathcal{M}_A$ and $\mathcal{M}_B$. The modular matrix $S'$ in such a system deviates from the bulk $S$ and can be explicitly computed using the techniques in \cite{kong2014anyon}.

\begin{figure}[h!]
\centering
\begin{tikzpicture}
% Disk with boundary segments A and B
\draw[fill=blue!10] (0,0) circle(2);
\draw[thick] (-2,0) arc[start angle=180,end angle=90,radius=2];
\draw[thick] (2,0) arc[start angle=0,end angle=270,radius=2];
\node at (-1,1.2) {$A$};
\node at (1,-1.2) {$B$};
\draw[->, thick] (-0.5,1.5) -- (-0.5,0.5);
\draw[->, thick] (0.5,-1.5) -- (0.5,-0.5);
\end{tikzpicture}
\caption{Schematic of a disk with boundary conditions $A$ and $B$. The junction of the two boundary segments reflects the fusion of module categories $\mathcal{M}_A$ and $\mathcal{M}_B$.}
\end{figure}
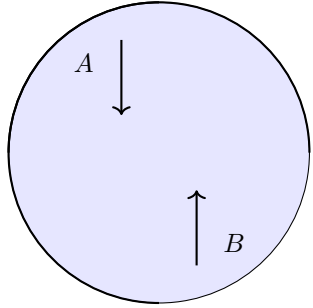

The fusion algebra associated with the boundary conditions $A$ and $B$ can be written as:
\begin{equation}
N_{ab}^c = \sum_{x} \frac{S_{ax} S_{bx} S_{cx}^*}{S_{0x}},
\end{equation}
where the sum runs over the fusion channels $x$ and $S_{ax}$ are elements of the modular matrix for the boundary conditions. This fusion algebra modifies the bulk matrix $S$, leading to a new matrix $S'$ given by:
\begin{equation}
S' = \begin{pmatrix}
S_A & \cdots \\
\vdots & S_B
\end{pmatrix}
\end{equation}
where \(S_A\) and \(S_B\) are the modular matrices corresponding to the boundary conditions \(A\) and \(B\), respectively. The size of \(S'\) is determined by the sum of the dimensions of the module categories associated with the boundary conditions. As the number of boundary conditions or their complexity increases, the structure of \(S'\) becomes more intricate.

In the simplest case, when \(A\) and \(B\) correspond to module categories with dimensions \(d_A\) and \(d_B\), the size of \(S'\) will be \((d_A + d_B) \times (d_A + d_B)\), reflecting the fusion of the two module categories.

The computation of the new modular matrix $S'$ can be done by applying the Verlinde formula:
\begin{equation}
N_{ab}^c = \sum_{x} \frac{S_{ax} S_{bx} S_{cx}^*}{S_{0x}}.
\end{equation}
This formula provides the elements of the new modular matrix $S'$, taking into account the mixing of module categories $\mathcal{M}_A$ and $\mathcal{M}_B$ at the junction.

\subsubsection{Example 2: Symmetry Defect Lines}
Defect lines preserving a global symmetry $G$ modify the path integral by inserting a $G$-twist, which can be captured algebraically via a $G$-crossed extension of the MTC \cite{barkeshli2019symmetry}. The modular transformations in this case are block-diagonal, with each block corresponding to a different twisted sector. This leads to a defect-modified modular matrix $S^G$ satisfying
\begin{equation}
S^G = \bigoplus_{g \in G} M_g S,
\end{equation}
where $M_g$ accounts for the defect's effect in the $g$-twisted sector.

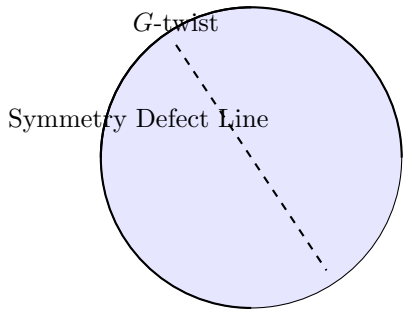
\begin{figure}[h!]
\centering
\begin{tikzpicture}
% Disk with a symmetry defect line
\draw[fill=blue!10] (0,0) circle(2);
\draw[thick] (-2,0) arc[start angle=180,end angle=90,radius=2];
\draw[thick] (2,0) arc[start angle=0,end angle=270,radius=2];
\draw[thick, dashed] (-1,1.5) -- (1,-1.5);
\node at (-1.5,0.5) {Symmetry Defect Line};
\node at (-1,1.8) {$G$-twist};
\end{tikzpicture}
\caption{Schematic of a disk with a symmetry defect line. The defect line modifies the path integral and is described by a $G$-twisted sector.}
\end{figure}

\paragraph{Mathematical Structure of \( S^G \)}
\paragraph{}
To understand how the modular matrix \( S^G \) is modified by the presence of symmetry defect lines. First note that defect lines introduce twisted sectors into the theory. Each element \( g \in G \) of the symmetry group generates a twisted sector, and the modular matrix \( S^G \) takes into account the effect of these sectors.

For each sector corresponding to \( g \in G \), the modular matrix is modified by a factor \( M_g \), which reflects the action of the defect line on the modular data. Thus, \( S^G \) is a block diagonal matrix where each block corresponds to a twisted sector indexed by \( g \in G \). The overall structure is given by

\[
S^G = \bigoplus_{g \in G} M_g S,
\]

where \( S \) is the bulk modular matrix, and \( M_g \) is a matrix that captures the defect’s effect in the twisted sector \( g \).

\paragraph{Twisted Sector Modifications}
\paragraph{}
The modification of the modular matrix in each twisted sector can be understood as follows. In the absence of defects, the modular matrix \( S \) captures the braiding and fusion properties of the system. However, when a defect line is present, the twisted sectors introduce new interactions between the anyons that depend on the symmetry group \( G \).

The matrix \( M_g \) accounts for these interactions by modifying the modular data in each twisted sector. It is related to the representation theory of the symmetry group \( G \) and the way it acts on the anyons. Specifically, \( M_g \) can be understood as a modification of the fusion and braiding statistics in the twisted sector corresponding to the element \( g \in G \).

\subsection{Categorical Description of Defects and Modular Data Transformations}

Defect lines in topological phases are categorically described by bimodule categories between different modular tensor categories (MTCs), or by braided tensor autoequivalences when the defects preserve the bulk topological order \cite{etingof2016tensor}. 

\subsubsection{Definition}
Given an MTC $\mathcal{C}$, a \emph{topological symmetry defect} is represented by a braided tensor autoequivalence $F: \mathcal{C} \to \mathcal{C}$. A \emph{domain wall} between two MTCs $\mathcal{C}, \mathcal{D}$ is modeled by a $\mathcal{C}$-$\mathcal{D}$ bimodule category.

\subsubsection{Theorem:ENO Aut Equivalence Correspondence}
Let $\mathcal{C}$ be a modular tensor category. Then the group of invertible topological defects in a $(2+1)$D TQFT based on $\mathcal{C}$ is isomorphic to the group of braided tensor autoequivalences\cite{etingof2016tensor}:
\[
\mathrm{Aut}^{\otimes, \mathrm{br}}(\mathcal{C}) \cong \text{Topological symmetry defects in } \mathcal{C}.
\]

In this framework, the action of the defect line on the modular data is induced by the pullback of the functor $F$, giving rise to a transformation:
\begin{equation}
S_{ij} \mapsto S_{F(i) F(j)}, \qquad T_i \mapsto T_{F(i)}.
\end{equation}
This transformation reflects how the statistical properties of anyons are modified in the presence of symmetry defects.

\paragraph{Remark}
\paragraph{}
The consistency of these transformations is constrained by higher-categorical coherence conditions. In particular, the $G$-crossed extension of an MTC requires a $G$-graded fusion category $\mathcal{C} = \bigoplus_{g \in G} \mathcal{C}_g$, equipped with $G$-actions and twisted associators satisfying generalized pentagon and hexagon axioms \cite{barkeshli2019symmetry}.

\subsubsection{Theorem:Module Category Description of Boundaries and Defects}
\paragraph{}
Let $\mathcal{C}$ be a modular tensor category and $A$ a special symmetric Frobenius algebra object in $\mathcal{C}$. Then the category of $A$-modules, $\mathcal{M}_A$, forms a module category over $\mathcal{C}$ and encodes the topological boundary condition associated with $A$ \cite{kong2014anyon}.

More generally, bimodule categories $\mathcal{M}_{A,B}$ over Frobenius algebras $A, B \in \mathcal{C}$ describe the Hilbert space at junctions between different boundaries, and modify the modular $S$-matrix into a boundary-sensitive form $S'$ that reflects fusion across domain walls:
\begin{equation}
(S')_{ij} = \sum_k \frac{S_{ik} \, (M_A)_{kj} (M_B)^*_{kj}}{S_{0k}},
\end{equation}
where $M_A$, $M_B$ are matrices encoding the action of boundary conditions $A$ and $B$ on the simple objects $k$ of $\mathcal{C}$.

\paragraph{Further Developments.}
These categorical structures are deeply connected to the Witt group of MTCs \cite{davydov2013witt}, which classifies modular categories up to Morita equivalence, reflecting the physical indistinguishability of TQFTs modulo boundaries and defects.

\section{Edge CFT Correspondence}

\subsection{Bulk-Edge Correspondence and WZW Models}

Topological phases of matter described by Chern-Simons theories in the bulk are holographically related to chiral conformal field theories on the boundary. In the case of $SU(N)_k$ Chern-Simons theory, the edge theory is a chiral Wess-Zumino-Witten (WZW) model with symmetry algebra $\widehat{\mathfrak{su}}(N)_k$.

\begin{itemize}
  \item The anyons in the bulk correspond to integrable representations of the affine Lie algebra $\widehat{\mathfrak{su}}(N)_k$ on the boundary.
  \item The modular $S$ and $T$ matrices of the topological phase match the modular data of the associated WZW model.
\end{itemize}

\subsection{Anomaly and Central Charge Matching}

The central charge of the edge WZW theory matches the gravitational Chern-Simons anomaly of the bulk theory:
\begin{equation}
c = \frac{k \cdot \text{dim}(\mathfrak{g})}{k + h^\vee},
\end{equation}
where $h^\vee$ is the dual Coxeter number of $\mathfrak{g} = \mathfrak{su}(N)$.

This matching ensures consistency between the chiral anomaly inflow from the bulk and the conformal anomaly on the edge. It also constrains the possible boundary conditions and defect insertions in the topological theory.

\subsection{Boundary Algebras and Fusion Rules}

At the boundary, the allowed boundary condition algebras form commutative Frobenius algebras within the representation category $\text{Rep}(\widehat{\mathfrak{g}}_k)$. These encode allowed fusion processes for edge excitations and reflect truncated fusion consistent with topological constraints.

Let $A$ be such a Frobenius algebra object. Then the open sector Hilbert space corresponds to the $A$-modules, and the open string fusion is controlled by the algebra structure of $A$. The consistency of boundary conditions and modular invariance require the Cardy condition to be satisfied.

\subsection{Examples}

\paragraph{Example 1: \boldmath$SU(2)_k$}

The primary fields are labeled by spins $j = 0, \frac{1}{2}, 1, \ldots, \frac{k}{2}$.

\begin{itemize}
  \item Central charge: $c = \frac{3k}{k+2}$.
  \item Modular $S$ matrix:
    \[
    S_{j_1 j_2} = \sqrt{\frac{2}{k+2}} \sin \left( \frac{(2j_1 + 1)(2j_2 + 1)\pi}{k+2} \right).
    \]
  \item Boundary algebras can be described using $A = \bigoplus_{j \in J} V_j$, where $J$ is a self-fusion-closed set of representations.
\end{itemize}

\paragraph{Example 2: \boldmath$SU(3)_2$}

This theory has 6 primary fields corresponding to the weights:
\[
(0,0), \ (1,0), \ (0,1), \ (1,1), \ (2,0), \ (0,2).
\]

\begin{itemize}
  \item Central charge: $c = \frac{16}{5}$.
  \item The modular $S$ matrix can be computed from affine $\mathfrak{su}(3)$ characters.
  \item Fusion rules exhibit triality symmetry.
  \item Boundary algebras correspond to Frobenius algebras that select subsets closed under fusion, e.g., $A = V_{(0,0)} \oplus V_{(1,1)}$.
\end{itemize}

\paragraph{Example 3: \boldmath$SU(4)_1$}

The primaries correspond to the weights:
\[
(0,0,0), \ (1,0,0), \ (0,1,0), \ (0,0,1),
\]
with abelian fusion.

\begin{itemize}
  \item Central charge: $c = 3$.
  \item Fusion corresponds to the $\mathbb{Z}_4$ group.
  \item Boundary algebras arise from full extensions corresponding to $\mathbb{Z}_2$ or $\mathbb{Z}_4$ subgroup condensations.
\end{itemize}

\subsection{Summary}

The edge CFT not only encodes the boundary excitations but also serves as a complete holographic dual of the bulk TQFT. The matching of modular data, central charge, and fusion categories across the bulk-boundary interface provides powerful constraints and tools for analyzing topological phases, especially in the presence of defects and domain walls.

\section{Conclusion}

This work provides a comprehensive analysis of boundary and defect effects on the modular data in SU$(N)_k$ Chern-Simons theories, with a focus on their role in modifying fusion rules and braiding statistics. By employing the framework of modular tensor categories (MTCs) and Frobenius algebra objects, we have demonstrated how different boundary conditions and symmetry defect lines can lead to deformations in the bulk modular matrices, $S$ and $T$, which encode the fusion and braiding properties of anyons.

Key contributions of this study include the explicit computation of the modified $S$-matrix, $S'$, in the presence of heterogeneous boundary conditions and defects, and its connection to the bulk modular data through the Verlinde formula. We have also explored the categorical structures underlying these modifications, specifically focusing on the functorial actions of defects and boundary conditions, and the pullback transformations that relate the modified modular matrices to their bulk counterparts. These results provide a unified perspective on the interplay between topological phases, boundary conditions, and defect lines.

In particular, we have applied this framework to the cases of SU$(2)_k$, SU$(3)_2$, and SU$(4)_1$, demonstrating how the fusion algebras at junctions between different boundary conditions can be derived and the resulting modular data deformations computed. Additionally, we have analyzed symmetry defects and their impact on the modular transformations, revealing the connection between global symmetries and twisted sectors in the MTC.

The work presented here opens several avenues for future research. In particular, the results pave the way for the systematic study of boundary conditions and defects in higher-rank Chern-Simons theories, as well as in non-unitary modular categories. Further exploration of categorical anomaly inflow and central charge matching across boundary and defect sectors will be crucial for a deeper understanding of the edge-bulk duality in topological quantum field theories. Moreover, the connection to topological quantum computation and holographic duality may yield further insights into the structure of low-dimensional quantum field theories and their applications in quantum information science.

The framework developed here also provides new insights into the relationship between modular data and the representation theory of quantum groups, affine Lie algebras, and higher categorical structures. As the study of defects and boundary conditions continues to evolve, it will be important to extend these ideas to more complex systems, including those with higher genus, curved backgrounds, and non-trivial spacetime topologies.

\appendix

\section{Mathematical Foundations of Defects and Boundaries in Modular Tensor Categories}

\subsection{Theorem: Invertible Defects and Braided Autoequivalences}
Let $\mathcal{C}$ be a modular tensor category (MTC). Then the group of invertible topological defects in $\mathcal{C}$ is isomorphic to the group of braided tensor autoequivalences of $\mathcal{C}$\cite{etingof2016tensor}:
\begin{equation}
\mathrm{InvDef}(\mathcal{C}) \cong \mathrm{Aut}^{\otimes,\mathrm{br}}(\mathcal{C}).
\end{equation}

\paragraph{Sketch of Proof:}
The proof relies on the following facts:
\begin{itemize}
    \item The invertible defects correspond to bimodule categories over $\mathcal{C}$ with certain rigidity and dualizability properties.
    \item The center $\mathcal{Z}(\mathcal{C})$ encodes the braided structure of $\mathcal{C}$.
    \item The Morita equivalence classes of such bimodules are in bijection with the braided autoequivalences of $\mathcal{C}$.
\end{itemize}

Formally, given an autoequivalence $F: \mathcal{C} \to \mathcal{C}$ preserving the braiding, one can construct an invertible bimodule category where the left action is the identity and the right action is twisted by $F$. This assignment is reversible, yielding the desired isomorphism.

\subsection{Theorem: Frobenius Algebras and Boundary Conditions}
Let $\mathcal{C}$ be a modular tensor category, and let $A$ be a symmetric special Frobenius algebra object in $\mathcal{C}$. Then the category of $A$-modules, denoted $\mathcal{M}_A = \mathcal{C}_A$, forms a module category over $\mathcal{C}$, describing a consistent boundary condition in the corresponding TQFT\cite{kong2014anyon}.

\paragraph{Sketch of Proof:}
\begin{itemize}
    \item The Frobenius algebra $A$ determines a condensation functor:
    \begin{equation}
        F_A: \mathcal{C} \to \mathcal{C}_A, \quad V \mapsto A \otimes V.
    \end{equation}
    \item The category $\mathcal{C}_A$ inherits a module structure via the monoidal structure of $\mathcal{C}$.
    \item The associativity and unital conditions of $A$ ensure that $\mathcal{C}_A$ is a module category.
    \item This construction respects fusion rules and braiding via modified $S$- and $T$-matrices, often written as $S^A$, $T^A$.
\end{itemize}

\paragraph{Example:}
In $SU(2)_k$ Chern-Simons theory, condensable algebras correspond to simple current extensions, and $A$ typically decomposes as a sum of objects closed under fusion. The boundary condition arising from such an $A$ corresponds to a gapped boundary in the physical theory.

\paragraph{Remark:}
The resulting module category encodes the Hilbert space associated to a boundary, and the modified Verlinde formula holds:
\begin{equation}
    N_{ab}^c = \sum_x \frac{S_{ax}^A S_{bx}^A (S^{-1})_{xc}^A}{S_{0x}^A},
\end{equation}
where $S^A$ is the modular matrix modified by the presence of $A$.

\printbibliography
\end{document}